# *Selective* Gas Sensing with a Single Pristine Graphene Transistor

Sergey Rumyantsev[1,2], Guanxiong Liu,[3] Michael S. Shur,[1] Radislav A. Potyrailo[4] and Alexander A. Balandin[3,5]

[1]Center for Integrated Electronics and Department of Electrical, Computer and Systems Engineering, Rensselaer Polytechnic Institute, Troy, New York 12180 USA

[2]Ioffe Physical-Technical Institute, The Russian Academy of Sciences, St. Petersburg, 194021 Russia

[3]Nano-Device Laboratory, Department of Electrical Engineering, Bourns College of Engineering, University of California – Riverside, Riverside, California 92521 USA

[4]Chemistry and Chemical Engineering, GE Global Research, Niskayuna, NY 12309 USA

[5]Materials Science and Engineering Program, University of California – Riverside, Riverside, California 92521 USA

**Abstract**

We show that vapors of different chemicals produce distinguishably different effects on the low-frequency noise spectra of graphene. It was found in a systematic study that some gases change the electrical resistance of graphene devices without changing their low-frequency noise spectra while other gases modify the noise spectra by inducing Lorentzian components with distinctive features. The characteristic frequency $f_c$ of the Lorentzian noise bulges in graphene devices is different for different chemicals and varies from $f_c$=10 – 20 Hz to $f_c$=1300 – 1600 Hz for tetrahydrofuran and chloroform vapors, respectively. The obtained results indicate that the low-frequency noise in combination with other sensing parameters can allow one to achieve the *selective* gas sensing with a *single* pristine graphene transistor. Our method of gas sensing with graphene does not require graphene surface functionalization or fabrication of an array of the devices with each tuned to a certain chemical.

Graphene – a planar sheet of carbon atoms arranged in honeycomb lattice – attracted a lot of attention owing to its extremely high mobility [1-4], thermal conductivity [5-6] and strongly tunable electrical conduction, which can be controlled with the gate bias [4]. Numerous device applications of graphene for high-frequency, analog, mixed signal communication and THz generation have been proposed [7-10]. Recent progress in graphene chemical vapor deposition (CVD) growth [11-12] and other synthesis techniques [13-14] together with development of the large-scale quality control methods for graphene [15] make practical applications of graphene feasible.

Graphene, with its extremely high surface-to-volume ratio, can become a natural choice material for sensor applications. The ultimate single-molecule sensitivity of graphene devices has been demonstrated at the early stages of graphene research [16]. It was suggested that the exceptional surface-to-volume ratio, high electrical conductivity, low thermal and $1/f$ noise [16-17], relatively low contact resistance [18-20], and ability to strongly tune the conductivity by the gate in graphene transistors make them promising for gas sensing applications [16]. Graphene resistivity, frequency of the surface acoustic waves (SAW), Hall resistivity, and the shift of the Dirac voltage have been used as sensing parameters [21-22]. The sensitivity of graphene devices to $NH_3$, $NO_2$, CO, $CO_2$, $O_2$, has been demonstrated. The high-gas sensitivity of graphene, which leads to its ability to detect ultra-low concentrations (down to <1 ppb) of different gases, and the linear dependence of the response to the gas concentration have been discussed in several publications (see reviews [21, 22] and reference therein). However, the selectivity of the graphene-based gas sensors is much less explored for the sensors utilizing all the above mentioned sensing parameters. In the present work, we demonstrate that the low-frequency noise can be used as the sensing parameter to enhance selectivity. We suggest that while the electrical resistivity or other DC parameter can serve as a quantitative parameter to measure the gas concentration, the low-frequency noise can allow one to discriminate between individual gases.

Owing to similarity of some properties between graphene and carbon nanotubes (CNTs), e.g. large surface-to-volume ration, high electron mobility, one can expect that graphene's potential for sensing can be extended to a wider range of applications following the CNT analogy. For example, CNTs have been used as nano-mechanical mass sensors with atomic resolution [23]. It has been demonstrated that a versatile class of nanoscale chemical sensors can be developed based on single-stranded DNA (ssDNA) for chemical recognition and CNT field-effect transistors (FETs) for the electronic read-out [24]. CNT FETs with ssDNA coating responded to vapors that caused no detectable conductivity change in bare devices [24]. An important observation made in experiments with ssDNA-decorated CNTs was that the sensor surface can be self-regenerating: the samples maintained a constant response with no need for sensor refreshing through at least 50 gas exposure cycles [25].

To improve the gas-response selectivity of graphene and related materials, several clever graphene preparation and functionalization methods have been developed. Reduced graphene oxide (RGO) platelets have also shown promise for vapor sensing [26]. The RGO films can reversibly and selectively detect chemically aggressive vapors such as $NO_2$ or $Cl_2$. The detection was achieved at room temperature (RT) for vapor concentrations ranging from 100 ppm to 500 ppb [26]. Two-dimensional "graphitic" platelets, oriented vertically on a substrate, have been shown to respond to relatively low concentrations of $NO_2$ and $NH_3$ gases [27]. Sensing applications of graphene were enabled not only via chemical but also biological functionalization, including by the use of phage displayed peptides [28] and DNA functionalization [29]. Several recent reviews summarized the state of the art of graphene gas sensors [21, 22, 30-33].

Sensor sensitivity is often limited by the electronic noise. Therefore, noise is usually considered as one of the main limiting factors for the detector operation. However, the electronic noise spectrum itself can be used as a sensing parameter increasing the sensor sensitivity and selectivity [34-36]. For example, exposure of a polymer thin-film resistor to different gases and vapors affects not only the resistance of the sensor but also the spectrum of the resistance fluctuations [34]. This means that by using noise as a sensing parameter in combination with the resistance measurements one can increase the sensor selectivity. This approach has been utilized for several types of gas sensors [35-36]. It is known that not only the amplitude but also the shape of the spectra changes under the gas exposure. In many cases, noise is a more sensitive parameter than the resistance. It has also been found that the changes in the resistance and noise are not always correlated and can be used as independent parameters in the analysis of the sensor response.

In this letter, we show that the low-frequency noise in graphene transistors is not always a detrimental phenomenon, which presents problems for its device application. We demonstrate experimentally that vapors of various chemicals affect the low-frequency noise spectra of graphene devices in distinctively different ways. Some vapors change the electrical resistance of graphene devices without changing their noise spectra while others introduce distinctive bulges over the smooth $1/f$ background. The characteristic frequencies of these bulges are clearly different for different chemicals. These unexpected findings demonstrate that noise can be used to discriminate between different gases. In combination with other sensing parameters this approach may allow to build a *selective* gas sensor with a *single* transistor made of *pristine* graphene which does not require an array of sensors functionalized for each chemical separately.

For the prove-of-concept demonstration, we adopted a standard mechanical exfoliation technique from the bulk highly oriented pyrolytic graphite [1-2]. The *p*-type highly-doped Si wafers covered with 300-nm thermally grown $SiO_2$ served as a substrate and back-gate for the graphene device channels. The single layer graphene (SLG) and bilayer graphene (BLG) samples were identified using the micro-Raman spectroscopy via deconvolution of the 2D band and comparison of the G peak and 2D band intensities. Details of our micro-Raman measurement procedures have been reported by some of us elsewhere [37-38]. The 10-nm Cr / 100-nm Au source and drain contacts were deposited on graphene by the electron beam evaporation (EBE). The bars connected graphene to the pre-deposited Cr/Au metal contact pads. Figure 1 shows scanning electron microscopy (SEM) images of several back-gated graphene transistors fabricated using the described approach.

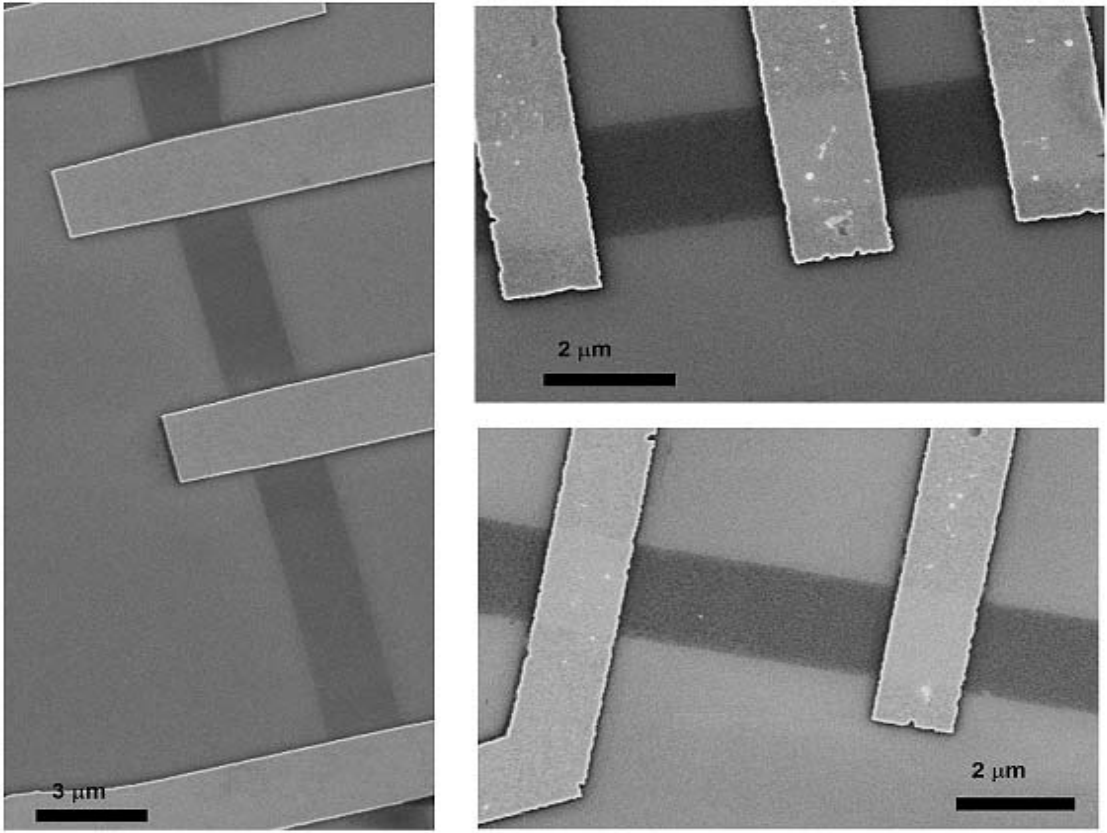

**Figure 1:** Scanning electron microscopy images of back-gated graphene devices used as selective gas sensors.

The low-frequency noise was measured in the common source configuration with a drain load resistor $R_L$=1-10 kΩ in a frequency range from 1 Hz to 50 kHz at room temperature (RT). The voltage-referred electrical current fluctuations $S_V$ from the load resistor $R_L$ connected in series with the drain were analyzed by a SR770 FFT spectrum analyzer. We have reported details of the noise measurements in graphene transistors in the ambient environment elsewhere [39-41]. For the present study, different vapors were generated by bubbling dry carrier gas (air) through a respective solvent and further diluting the gas flow with the dry carrier gas. In this way, all vapors were generated at concentrations of ~0.5 P/P$_o$, where P is the vapor pressure during the experiment and P$_o$ is the saturated vapor pressure. Upon completing the measurements with one vapor and before the exposure to another vapor, each device was kept in vacuum for several hours at RT.

Figure 2 shows a typical current voltage characteristic of a back-gated transistor with the SLG channel measured at ambient conditions. The charge neutrality point – also referred to as Dirac voltage – was about 10-20 V for the as fabricated devices selected for this study. The field-effect and effective mobilities extracted from the current-voltage characteristics were in the range 5000 -10000 $cm^2/Vs$. All devices revealed the hysteresis under the direct and reverse gate voltage scans. This is a well-known effect [42-44] attributed to the slow carrier relaxations due to the presence of deep traps. Our pulse measurements showed that these relaxation processes are non-exponential within the time scale from ~20 ms to at least 1000 s. In order to avoid this unstable behavior we performed all measurements at zero gate voltage, i.e. on the "hole" part of the current voltage characteristic (see Figure 2).

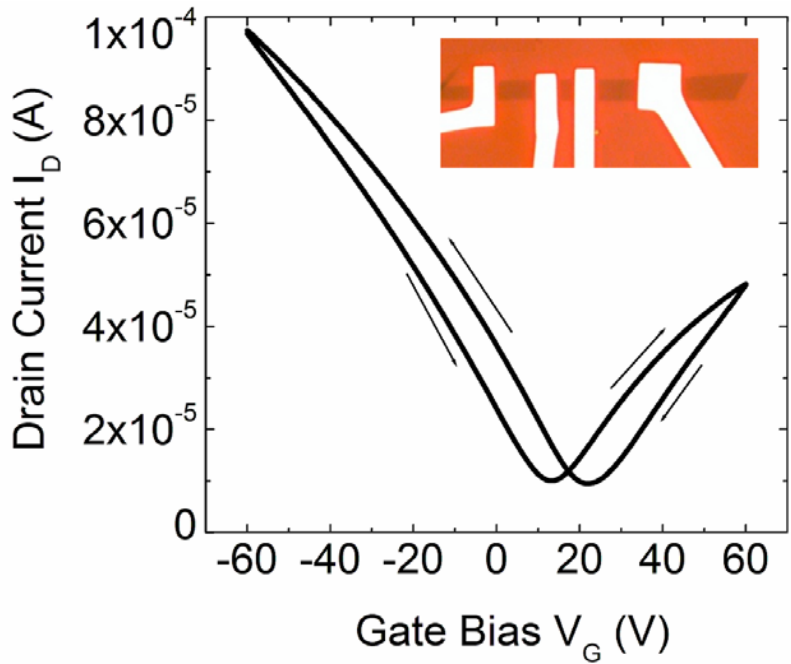

**Figure 2:** Transfer current-voltage characteristic of a typical back-gated graphene transistor used for the gas sensing tests. The arrows indicate the direction of the gate voltage sweep. The inset shows an optical microscopy image of the graphene transistor with the top metal electrodes.

After measuring the transistor current-voltage characteristics the devices were exposed to the laminar flow of individual vapors such as methanol, ethanol, tetrahydrofuran, chloroform, acetonitrile, toluene, and methylene chloride. An inset in Figure 3 shows an example of the resistance change under the influence of ethanol. As seen, the resistance response is rather slow taking several hundreds of seconds to reach the steady state condition. The process of degassing is even slower but can be accelerated by the exposure to ultraviolet (UV) light. In the inset, one of the arrows shows the moment of time when the 280-nm light-emitting diode (LED) was turned on. The effect of UV cleaning is known for carbon nanotubes and graphene gas sensors [45-46]. However, we found that extending exposure to UV can irreversibly alter the graphene device characteristics. Therefore, this method of degassing was not used in our selective gas sensing experiments.

Figure 3 presents examples of the noise spectra measured in open air and under the influence of tetrahydrofuran and acetonitrile vapors. The noise was measured in ~1 min after the device exposure to the vapor. The measurements were repeated several times with a time interval of ~ 5min. There are two and three overlapping spectra in Figure 3 for acetonitrile and tetrahydrofuran, respectively corresponding to multiple measurements indicating excellent reproducibility of the noise measurements. As a result of the vapor exposure the noise increases and the shape of the noise spectra changes. The appearance of characteristic bulges, over 1/$f$ noise background, indicates a contribution of the random processes with the well-defined relaxation time [47]. In the case of a single relaxation time the noise spectrum has the form of the Lorentzian:

$$S \propto \frac{1}{1+(\omega\tau)^2}, \tag{1}$$

where $\tau$ is the relaxation time and $\omega=2\pi f$ is the circular frequency.

In semiconductors this kind of excess noise is often associated with the generation-recombination (G-R) noise [47]. It is conventionally attributed to fluctuations of the occupancy of the local energy levels. The temperature dependence of the G-R noise in semiconductors allows one to determine all parameters of the given local level, which is the subject of the so-called noise spectroscopy [48]. Other mechanisms also can lead to the Lorentzian type of the spectra. Particularly, mobility fluctuations with a single relaxation time also reveal themselves as the Lorentzian bulges [49]. In addition to the Lorentzians observed due to the G-R or mobility fluctuation processes there have been reports of the Lorentzian noise induced by shot or Nyquist noise in MOSFETs [47]. In our previous studies of low-frequency noise in graphene devices we found that the number-of-carriers fluctuation mechanism, typically responsible for the GR noise, cannot explain the gate bias dependence of noise in graphene [17]. For this reason, we avoided using the tern GR noise in reference to the observed bulges in the low-frequency spectra of graphene devices exposed to vapors. Here and below we adopted the term *Lorentzian noise* instead.

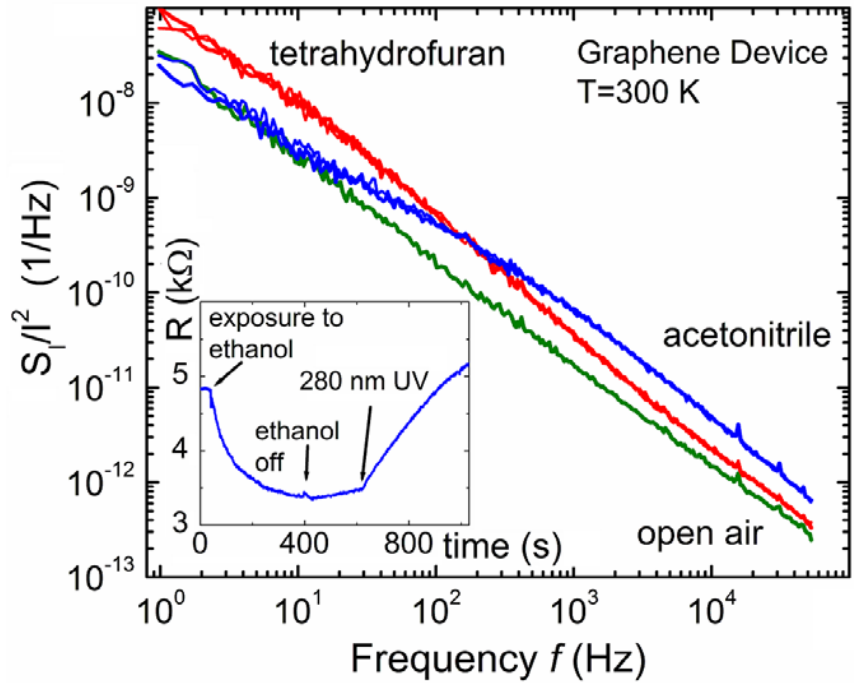

**Figure 3:** Noise spectra of SLG transistors measured in open air and under the exposure to acetonitrile and tetrohydrofuran vapors. The gate bias is $V_G=0$ V with the source-drain voltage is $V_D=100$ mV. The inset shows the resistance response of the graphene transistor to the exposure of ethanol as a function of time. The gate bias for the data presented in the insert is $V_G=0$ V.

In order to establish the characteristic frequency $f_c=1/2\pi\tau$ of the Lorentzian noise for each given vapor, in Figure 4, we plotted the noise spectra multiplied by the frequency $f$, i.e. $S_I/I^2 \times f$, versus $f$. As one can see, these dependencies have well distinguished maxima at frequencies $f_c$, which are different for different vapors. This result suggests that the frequency $f_c$ can be a distinctive signature of a given vapor.

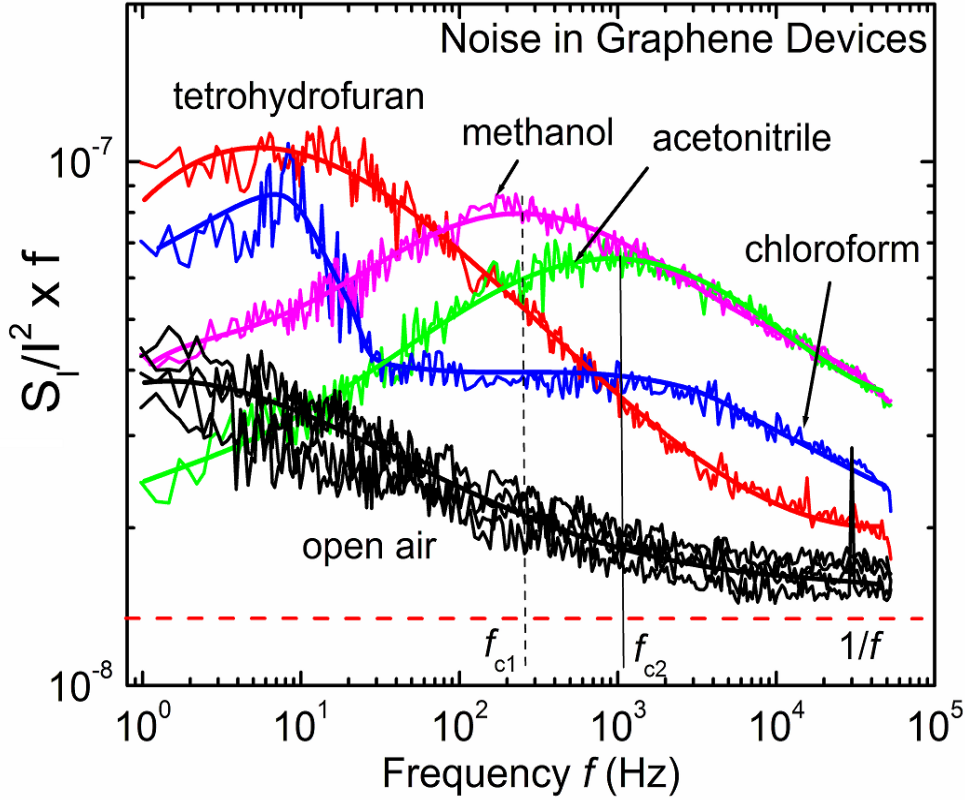

**Figure 4:** Noise spectral density $S_I/I^2$ multiplied by frequency $f$ versus frequency $f$ for the device in open air and under the influence of different vapors. Different vapors induce noise with different characteristic frequencies $f_c$. The frequencies, $f_c$, are shown explicitly for two different gases. The solid lines show the polynomial fitting of the experimental data. The difference in the frequency $f_c$ is sufficient for reliable identification of different gases with the same graphene transistor. For comparison the pure $1/f$ noise dependence is also indicated.

From the physics point of view, there can be two reasons for the Lorentzian noise in graphene appearing under the gas exposure. First, the gas molecules can create specific traps and scattering centers in graphene, which lead to either number of carriers fluctuation due to the fluctuations of traps occupancy or to the mobility fluctuations due to fluctuations of the scattering cross sections [47-50]. Another scenario is that the kinetic of the molecule adsorption and desorption contributes to noise. The characteristic time scale for the adsorption of vapors was several hundreds of seconds. It is even longer for the degassing. This corresponds to much lower characteristic frequencies than those observe in the present work. Therefore, we concluded that the appearance of the Lorentzian noise is related to the charge traps created as a result of vapor exposure. However, the specific mechanism of the observed Lorentzian noise in graphene can be different from that in semiconductor devices.

Table I presents the characteristic frequencies $f_c$ and the relative resistance $\Delta R/R$ changes in graphene devices for different vapors ($R$ is the resistance). In spite of the large resistance changes under exposure to toluene and methylene chloride the noise spectra did not alter under exposure to these vapors. One can see from Table I that a combination of the resistance change and frequency $f_c$ provides a unique characteristic for identification of the tested chemicals. The data summarized in Table I can be used for the selective gas sensing using a single graphene transistor. The latter is a major positive factor for sensor technology since it allows one to avoid fabrication of a dense array of sensors functionalized for individual gases.

We tested the selected set of chemicals vapors on different graphene device samples and alternated different vapors for the same samples. We found that our results were well reproducible provided that the graphene transistors were degassed by keeping in vacuum at RT for at least 2-3 hours prior the measurements. Figure 5 shows the $S_I/I^2 \times f$ versus frequency $f$ dependencies for three different graphene transistors under exposure to the acetonitrile vapor. As one can see despite different amplitude of the noise the frequency $f_c$ is the same for all three devices.

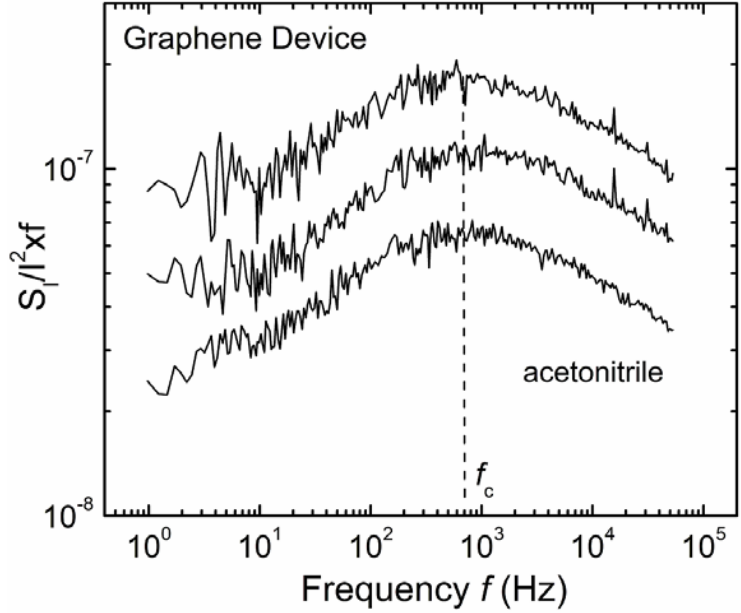

**Figure 5:** Noise spectral density $S_I/I^2$ multiplied by frequency $f$ versus frequency $f$ for three different single-layer-graphene transistors exposed to acetonitrile vapor. Note the excellent reproducibility of the noise response of the graphene devices showing the same frequency $f_c$ for all three devices.

**Table I: Frequency $f_c$ and $\Delta R/R$ in Graphene for Different Vapors**

| Vapor | $f_c$ (Hz) | $\Delta R/R$ % |
|---|---|---|
| Ethanol | 400-500 | -50 |
| Methanol | 250-400 | -40 |
| Tetrahydrofuran | 10-20 | +18 |
| Chloroform | 7-9 and 1300-1600 | -25 |
| Acetonitrile | 500-700 | -35 |
| Toluene | NA | +15 |
| Methylene Chloride | NA | -48 |

In conclusion, we found that chemical vapors change the noise spectra of graphene transistors. The noise spectra in open air are close to the *1/f* noise. Most vapors introduce Lorentzian bulges with different characteristic frequencies $f_c$. The frequency $f_c$ of the vapor-induced Lorentzian noise and the relative resistance change *ΔR/R* serve as distinctive signatures for specific vapors enabling highly selective gas sensing with a single graphene device. The noise spectra are well reproducible and can be used for reliable chemical sensing. The observation of the Lorentzian components in the vapor-exposed graphene can help in developing an accurate theoretical description of the noise mechanism in graphene.

**Acknowledgements**

The work at RPI was supported by the US NSF under the auspices of I/UCRC "CONNECTION ONE" and by the NSF EAGER program. SLR acknowledges partial support from the Russian Fund for Basic Research (RFBR) grant 11-02-00013. The work at UCR was supported by the Semiconductor Research Corporation (SRC) and the Defense Advanced Research Project Agency (DARPA) through FCRP Center on Functional Engineered Nano Architectonics (FENA), DARPA Defense Microelectronics Activity (DMEA) and the US National Science Foundation (NSF). The work at GE was supported by GE Corporate long term research funds. This publication was made possible, in part, by NPRP grant # NPRP 09-1211-2-475 to RPI from the Qatar National Research Fund. The statements made herein are solely the responsibility of the authors.